# Impact of COVID-19 on human mobility and retail sales in the US

Ayobami Esther Olanrewaju[1] & Patrick E. McSharry[1,2,3]

[1] Carnegie Mellon University Africa, Kigali, Rwanda
[2] African Centre of Excellence in Data Science, University of Rwanda, Kigali, Rwanda
[3] Oxford MAN Institute of Quantitative Finance, University of Oxford, Oxford, UK
mcsharry@cmu.edu,  aolanrew@andrew.cmu.edu

## Abstract
Due to the COVID-19 pandemic, governments had to rapidly implement lockdown policies that restricted human mobility to suppress the spread of the disease and reduce mortality. Although this intervention was successful in saving lives, it delivered a serious shock to the economy and decimated retail sales because of travel restrictions. In 2020, the United States (US) experienced the biggest fall in annual GDP since 1946 at -3.5%, plummeting by as much as -31.2% in the second quarter. The US retail sector accounts for 6% of the US GDP and 6.3% of the workforce. Because of the movement restrictions resulting from government responses to the pandemic, US retail sales declined by -22% in April 2020 compared to the previous year. This study looks at the stringency of government policies, mobility patterns and implied compliance levels. The relationships between these variables and the influence on retail sales serves to understand past human behavior and to better prepare for future pandemics. Retail losses varied dramatically across the US states, from -1.6% in Mississippi to -38.9% in Hawaii. Geography is a key determinant, with states in the west and northeast being most affected, while those in the south were relatively resilient. Regression was used to identify statistically significant state-level characteristics. The greatest losses occurred in states with a high percentage of Democrat voters in the 2020 Presidential Election and also those with large populations. A ten percentage increase in the Democrat vote is associated with a 2.4% increase in retail sales loss. States with a high percentage of adults with less than a high school diploma were most resilient. The number of trips of less than one-mile per capita is defined as the mobility index as it has the greatest influence on retail sales, on average, across the US states. An increase of 10% in this mobility index is associated with a 4.6% increase in retail sales. All states were generally compliant and exhibited a reduction in mobility with increasing stringency. A rise of 1% in the stringency index is associated with a decline of 1% in the mobility index. States with a high percentage of Democrat voters,  large populations, and located in the west tend to be most compliant. A 10% rise in the proportion of people voting Democrat is associated with a 5% increase in compliance. The dependency of retail on mobility is particularly high in states with a low percentage of Democrat voters.

# Impact of COVID-19 on human mobility and retail sales in the US

## 1. Introduction

This introduction section consists of the background, motivation, objectives, literature review, contribution, and layout of the research.

**1.1 Background**:

Coronavirus (COVID-19) is a respiratory and infectious disease named Severe Acute Respiratory Syndrome Coronavirus -2 (SARS-CoV-2) (Yuki, Fujiogi and Koutsogiannaki, 2020). There are several coronavirus variants of concern: Alpha, Beta, Gamma, Delta, and Omicron (Velavan and Meyer, 2020; Duong, 2021). Since being initially detected in late 2019, the disease has mutated, leading to some variants emerging and others eventually disappearing. For example, the Delta variant was detected in October 2020 and the Omicron variant found in different countries in November 2021 (WHO, 2022). Mutations cause changes in the spread of the virus, the severity of the disease, the performance of vaccines, social responses, and diagnosis without changes in the virus properties (WHO, 2022). The World Health Organization (WHO) declared COVID-19 a public health emergency on January 30th, 2020, and a global pandemic on March 11th (WHO, 2020a; WHO, 2020b).

The number of COVID-19 cases worldwide has now exceeded half a billion and cases have been recorded in almost every country across the globe. The death toll has surpassed six million, with higher mortality recorded in the elderly population and mild symptoms in children (Yuki, Fujiogi and Koutsogiannaki, 2020; Fauci, Lane and Redfield, 2020). The swift spread of the disease among people and the worldwide outbreak have impacted how we live, modified our interactions and sparked fears of impending economic crises and recession (Nicola et al., 2020; Josephson, Kilic and Michler, 2021). The disease exacerbated income, occupational, and ethnic inequality (Barlow and Vodenska, 2021; Etienne, 2022; Sekhri Feachem, Sanders and Barker, 2021; Paul, Englert and Varga, 2021; Ferreira, 2021). Lower-skilled workers and the uneducated were the worst affected (Paul, Englert and Varga, 2021). Richer countries experienced a more significant economic contraction than poorer ones, and places with ethnic disparities and stereotyping of race led to challenges in implementing mitigating procedures, which allowed more diseases to spread (Nicola et al., 2020; Barlow and Vodenska, 2021). Since its outbreak, different countries in America, Europe, Asia, and others have been affected. The United States (US) COVID-19 cases and deaths have been among the highest globally. The US accounts for 25% of the global COVID-19 cases and 20% of the deaths worldwide, even though the US only makes up 4% of the world population (Sekhri Feachem, Sanders and Barker, 2021).

The rapid spread of the disease has necessitated the government's prompt action to initiate a restriction of movement and lockdowns to curtail the transmission of the disease, allowing public health agencies to design new interventions, and possibly reduce the need for future lockdowns (Sekhri Feachem, Sanders and Barker, 2021). Aside from movement restrictions, different spread reduction modes have been practiced, such as wearing face-coverings, quarantine, isolation of infected people, social distancing, and regular cleaning of surfaces (Bachman, 2020). Because of the restrictions resulting from the disease, different facets of the economy were disrupted. The virus affected every sector; however, healthcare, transport, tourism, manufacturing, aviation, education, small-medium size retail, and food sectors were most affected (Fauci, Lane and Redfield, 2020; Nicola et al., 2020; Paul, Englert and Varga, 2021; Rio-Chanona et al., 2020). Schools and workplaces were locked down. Restaurants, services, tourism, manufacturing, and other businesses suffered because movements were restricted and social distancing was enforced. The most affected people were in low-wage occupations, while those with high wages and remote jobs were relatively immune to being confined to their homes.

Furthermore, specific sectors like healthcare expanded because of the need to find vaccines and treat infected people. At the end of 2020, almost 27 million Americans became underemployed, unemployed, or dropped out of the workforce (Sekhri Feachem, Sanders and Barker, 2021). Since the American healthcare system relies on an employment-based private healthcare insurance model, the loss of jobs means the people affected may have lost their healthcare coverage during the pandemic (Sekhri Feachem, Sanders and Barker, 2021; Bergquist, Otten and Sarich, 2020). A loss of $11.4 trillion over the next decade has been attributed to COVID-19 and an additional loss due to premature death and long-term physical and mental impairment, thereby raising the losses to between $25 to $30 trillion or 135% of annual GDP (Rio-Chanona et al., 2020). Aside from the losses during the pandemic, food insecurity doubled and almost tripled for families with children (Schanzenbach and Pitts, 2020).

In this paper, we evaluate the impact of COVID-19, looking specifically at the retail sector of the United States. Using publicly available datasets, we also assess people's mobility in relation to different lockdown measures associated with the virus and quantify the financial impact and variability across the US states. By understanding the past behavior of the population and compliance levels in particular, we hope to be better prepared for future pandemics.

**1.2 Motivation**:
COVID-19 has captivated the world since its outbreak in 2019 and influenced many important decisions. It has disrupted all economic sectors such as manufacturing, health, education, agriculture, tourism, etc. The retail industry is a vital sector of an economy, and it relies on the movement of people and their ability to purchase goods and services. For the US, the retail sector accounts for 6.3% of the US employment and 6% of the Gross Domestic Product (GDP). Given that US healthcare access is intrinsically linked to work, losing a job often means losing medical insurance, which can later have a negative impact on livelihoods and economic productivity. Also, fewer purchases implies financial losses for the retail sector, which will lower the country's GDP.

**1.3 Objectives**:
This research addresses the following objectives
1. To quantify the financial loss of the retail sector due to COVID-19
2. To measure variability in impact by the US states

3. To explain the difference in effects using state-level explanatory variables

**1.4 Literature review**:
COVID-19 is a public health challenge that has affected how we live our lives and poses a threat to human development and well-being. It increased poverty, altered consumption patterns, created market anomalies, led to revenue and life loss, and further widened the inequality gap (Baldwin and Weder di Mauro, 2020; Paul, Englert, and Varga, 2021; Bartik et al., 2020). It also exposed racial bias, health inequalities, and social injustice and led to a reset of values (Hatef et al., 2020). Desjardins (2020) and Assefa et al., (2022) opined that the propagation of the virus shows geographical trends as it spreads from the more dense areas to more rural parts and countries with a higher human development index have higher cases and deaths. Lockdowns and social distancing were strategies adopted by the government to suppress the disease. Hallas et al. (2021) used the Oxford COVID-19 Government Response Tracker (OxCGRT) indicator to determine different US government responses to the virus by region and political voting patterns. They discovered that home permanence tended to increase and visits to non-essential retail tended to decrease when policy stringency was highest, and these effects gradually decline after periods of high stringency, possibly indicating policy fatigue. While lockdowns worked as a temporary fix, they had a detrimental impact on production and thereby caused economic downturns (Brinca, Duarte and Faria-e-Castro, 2021; Gupta, Simon and Wing, 2020).

Looking at research on COVID-19 in the US and its impact, Rio-Chanona et al. (2020) studied supply and demand shocks by classifying industries as essential or nonessential and constructing a remote labor index that measures the ability of different occupations to work from home. Compared to pre-COVID-19, the shocks threaten 22% of the US economy's GDP, jeopardize 24% of jobs and reduce wages by 17%. However, the economic impact of COVID-19 has not been uniform. Paul et al. (2021), while applying a boosted decision tree on survey data to measure embedded correlations in the data and Shapley values, discovered that nonwhite women, the less educated, racial minorities, and low-income earners felt the effect of the virus the most. Barlow and Vodenska (2021) suggested that manufacturing, leisure, hospitality, and professional services were the most affected sectors. While some companies were hit hard and had to fire workers and cope with depleted sales, others benefited from a surge in sales and expanded their workforce (McKinsey & Company, 2020). Shopping behavior has had crucial implications on business operations, which have changed due to COVID-19 and may be long-lasting, thereby reforming the retail sector (Melo, 2020).

Much of the literature investigated socio-economic impacts by studying jobs, stringency, cases, deaths, and mobility separately. In contrast, this research looks at how COVID-19 led to government policy responses which decreased mobility and ultimately caused a drastic decline in retail sales. In addition, the study pays particular attention to the state-level characteristics that explain the variability across different US states. Therefore, by evaluating the financial impact of COVID-19 on the retail sector and its variability across the states, policymakers can learn how different responses restrict mobility and affect the US economy. Moreover, quantifying each of the relationships between policy and mobility and between mobility and retail sales makes it possible to be better prepared for future pandemics by striking a balance between suppressing the spread of disease and generating substantial economic losses.

**1.5 Contribution**:

This research extends the literature on the effects of COVID-19 in the US by focusing on how COVID-19 led to a decrease in movement due to lockdowns, how reduced mobility decimated the retail sector, and the resulting adverse financial impact on the US economy. The specific factors that reflect the link between mobility and retail sales are examined and the variation across US states is investigated. By quantifying these relationships, the research provides insights for US policymakers, economists, and researchers to make better informed decisions and evaluate the impact of the virus on the sector that accounts for approximately 6% of both the total US employment and US GDP (U.S. Bureau of Economic Analysis, 2021; US Census Bureau, 2020).

**1.6 Layout**:
This paper is divided into sections containing the methods, results, discussion, and conclusion. First, the methods section shows the case study data sources and the techniques employed to answer each of the research questions. Next, the results section describes the insights obtained from the analysis. The discussion section then examines the results obtained in detail. Finally, the conclusion section summarizes the findings, states the limitations of the work, and offers recommendations.

# 2. Methods

The methods section contains information about the study, available datasets, data sources, and data collection tools. It also includes information on data processing techniques used to transform data into features for further analysis. Finally, it describes the models and any visualization techniques used.

**2.1 Data Sources and Collection tools**
The research uses publicly available data from the United States Census Bureau (USCB), United States Bureau of Economic Analysis (BEA), Our World in Data (OWID) and Covid-19 Government Response Tracker (OxCGRT), United States Bureau of Transport Statistics (USBTS), The Cook Political Report, Federal Reserve Economic Data (FRED), the Organization for Economic Co-operation and Development (OECD), and the US Department of Agriculture's (USDA) Economic Research Service (BEA, 2020; BTS, 2022; OWID, 2022a; OWID, 2022b; OxCGRT, 2020; USCB, 2022; OECD, 1955; USDA, 2022; Wasserman et al., 2020 ). All the data are sampled monthly and consist of values for each of the 50 US states, the district of Columbia, and US national data. The data observation period of this project is January 2019 to December 2021. The selected variables and transformations applied are described in greater detail in Section 2.2. The following datasets and variables were used to support the analysis:

# Impact of COVID-19 on human mobility and retail sales in the US

1. **Monthly Retail Trade:** USCB's monthly retail trade data shows the year-on-year changes for retail transactions, excluding non-store retailers for all US states. The data is from January 2019 to December 2021, sampled monthly and recorded in percentages. The information incorporates survey data obtained via mail from retail businesses, administrative data, and third-party data.
2. **USBTS Mobility:** The USBTS's "trip by distance" data consists of movements that include a stay of longer than 10 minutes at an anonymized location far from home. This data captures all modes of transportation such as driving, rail, transit, and air. The monthly data ranges from January 2019 to December 2021, and is recorded in miles. The initial daily measurements are obtained from a mobile device data panel from combined sources that address geographic and temporal sample variation. These daily measurements are then aggregated to compute a monthly average. The variables include the population at home and that not at home. State-level counts are normalized by the relevant populations to create a per capita value for each state. Trips less than 10 miles are categorized as short trips, trips greater than or equal to 10 miles and less than 100 miles are called medium trips, and trips of 100 miles or more are denoted as long trips.
3. **Google Mobility**: Information about the mobility of individuals regarding retail and recreation and residential is obtained from Google Mobility trends, which is retrieved anonymously from Google Maps. The data is computed relative to the baseline days before the pandemic outbreak. The raw data is then smoothed using a seven-day rolling average to remove the intraweek seasonality.
4. **COVID data:** The COVID cases, deaths, and vaccination information is also accessed from OWID and based on data collected from John Hopkins University. This data is normalized by population, and again a seven-day rolling average is applied. The research extracted information for the US from the dataset. The variables used are new cases smoothed per million and new deaths smoothed per million
5. **Stringency Index:** The stringency index is computed by the OxCGRT to reflect the level of government response using publicly available sources such as news articles, government press conferences, and briefings. The index is a number between 0 and 100, and it contains information for containment and closure policies (C) and health system policies (H). It relies on eight (C1-C8) indicators and the single H1 indicator. The indicators are school closure, workplace closure, canceled public events, restrictions on gathering, closed public transport, stay-at-home requirements, restrictions on internal movement and international travel control, and public information campaigns. The data is recorded daily for each US state, but this research uses the stringency index value at the end of the month between January 2020 and December 2021.
6. **Economic Output:** The Gross Domestic Product (GDP) for each state in 2019 is obtained from BEA, which monitors economic output. The GDP is then normalized by the population to calculate a per capita value for each state. GDP per capita for the entire US from January 2019 to December 2021 is obtained from FRED.
7. **State-level Data**: This data is based on surveys undertaken by the USCB and was obtained from the USDA. Education Attainment is measured from 2015 to 2019. Population, land area and population density were provided by USDA. Median household income was available for 2019 by USDA.
8. **Election Voting Pattern:** The election voting patterns recorded by the Cook report provide data on the 2020 Presidential election between Trump and Biden. It gives the percentage of votes cast by democrats, republicans, and the political orientation of a state.

## 2.2 Data Processing Techniques
This research adopted the following data processing techniques to utilize the available data.

### 2.2.1 Data Preparation
Data cleaning involved removing duplicated data and date-time conversions which were necessary to obtain the required monthly retail sales values. The format of the temporal variables had to be converted to compare the different time series. It was then necessary to combine the data so that stringency, mobility, and retail could be compared and analyzed. Data normalization is crucial to ensure that the different states can be compared. The population normalized all the trip-by-distance data as it was divided by population to facilitate comparability across the US states. The US population value normalizes the state population variable. The stringency index ranges from 0 to 100 and year-on-year state retail sales are in percentages.

### 2.2.2 Feature Engineering
This process involves creating a new feature or transforming an old one into a more predictive feature that improves the model's quality. Label encoding helps transform categorical variables into an indicator variable that the machine learning algorithm understands, such as the Republican variable which is a dummy variable that represents 1 for Republicans and 0 for Democrats. A similar approach was used to encode the US regions.

### 2.2.3 Feature Selection
This is the process of picking the best features that are most relevant when constructing a model, with the aim of obtaining accurate predictions. Backward stepwise regression is used in this project for feature selection. The approach starts with all the candidate variables included and gradually removes unimportant features until arriving at those that best explain the data. A probability threshold of 0.05 is used to determine if a feature is statistically significant. A feature with a p-value above 0.05 is regarded as statistically insignificant and removed.

### 2.2.4 Statistical techniques and visualization methods
The research adopts line plots, scatterplots, bar charts, and tables to present the results. The table shows the summary statistics, while the charts help to visualize how the results vary across the US states. Scatterplots and correlations quantify relationships between variables. Backward stepwise regression extracts significant features with a p-value of less than 0.05. A linear model is then fit between the explanatory variables and the dependent variable. The p-values are categorized into three groups to determine the strength of significance of each variable: $p<0.05$ (low); $p<0.01$ (moderate); and $p<0.001$ (high). Finally, the model's coefficients, t-statistics, p-values, R-squared and adjusted R-squared values are presented in a summary table.

### 2.2.5 Modeling the economic impact
To better understand the sizes of the financial losses in the retail sector that arose from COVID-19, there are two important relationships to consider. First, it is necessary to quantify how individuals react to government policies and specifically what kind of mobility behavior is observed as the stringency level is increased. Second, while retail clearly suffered from lockdowns, the specific mobility measure that best predicts retail sales can be identified. This is achieved by using the Pearson correlation coefficient to compare the relationship between various mobility variables and retail sales. This analysis ultimately helps to determine which specific factors and therefore changes in human behavior contributed to the large losses in retail sales.

## 3. Results
This section presents the results of this research. After describing the trends and relationships between the variables at the US national level, the financial losses are estimated at the state level. The state-specific factors that influenced the size of the losses are identified. A measure of compliance is defined and estimated for each state. Finally the factors that explain compliance and the dependence of the retail sector on mobility are explored.

### 3.1 US National Situation
The headline graph (Figure 1) shows the monthly time series for each variable of interest to understand the motivation behind the study and to visualize the impact of COVID-19 on retail sales. Although focusing on the entire US, the chart provides a holistic view of the impact of COVID-19 and the interactions between different variables that will be further investigated. The spread of COVID-19 infections, consequent rising death toll and overrun hospitals motivated the government to respond with lockdowns and mobility restrictions. On average, the more severe the restrictions the greater the impact on retails sales. Concerted efforts to suppress COVID-19 in April 2020 effectively brought movement to a halt across the US. At this time, the OxCGRT stringency index (navy blue) had rapidly increased to reach 72% and the number of trips (gray) based on the BTS calculation of a person spending more than ten minutes at a location away from home declined by 40% from pre-pandemic levels. All mobility measures for trips, people away from home and those at retail & recreation plummeted in April 2020 at the height of the lockdowns and are still recovering. Similar patterns were observed using BTS data and Google Mobility data with larger numbers of people staying at home and less trips taking place. This dramatic result of the lockdown is illustrated by the increasing number of people staying at home (green) and rising Google Mobility Residential (olive) and a decrease in Google Mobility retail & recreation (blue). Unfortunately, the adverse consequences for the economy were immense with the year-on-year change in retail (black) dropping by more than 20% in April 2020. Over time with the roll out of vaccines in 2021, the stringency index gradually declined as people learned to live with the virus. While the population at home is decreasing, neither the mobility measure for retail & recreation nor the number of trips have returned to 2019 levels.

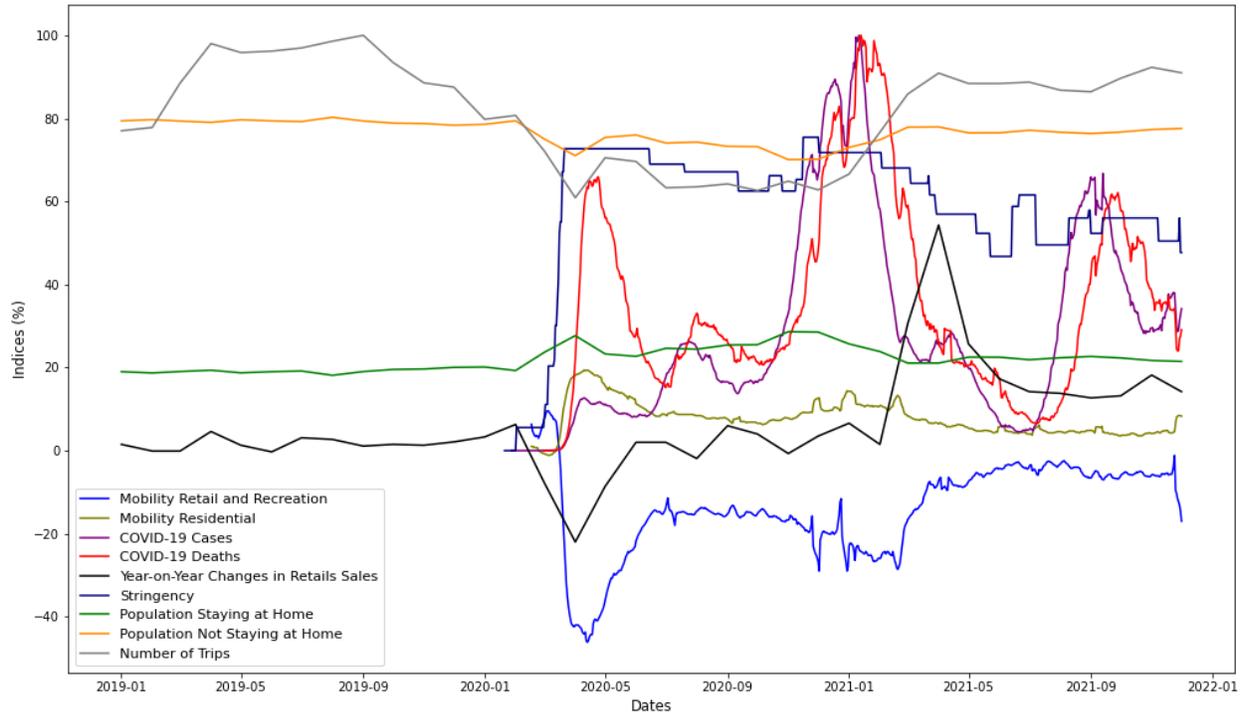

Figure 1: Headline graph showing COVID-19 cases, deaths, stringency index, mobility indices, and year-on-year changes in retail for the US from January 2019 to December 2021.

**3.2 Quantifying the financial losses of the retail sector due to COVID19**

Government responses to COVID-19 worldwide involved lockdowns and restrictions on mobility. The consequence was severe economic losses corresponding to a contraction of the US economy, reflected in GDP decreasing by 7.4% between April 2019 and April 2020. This loss of economic output was particularly strong in the retail sector due to the constraints on mobility and mandated closures. The greatest decline in retail sales in the US occurred in April 2020, falling by -22% compared to the previous year (Figure 1).

By investigating yearly changes in retail sales at the state level, the widespread losses are apparent (Figure 2). The peak in April 2021 is simply due to the apparent recovery from the immense fall that occurred twelve months previously. Many states (gray) follow a similar pattern and are close to the US average, suggesting the widespread impact of the pandemic. In contrast, a few states, such as Vermont, Mississippi, and Kentucky, stand out as being relatively resilient to the COVID-19 shock and displayed minimal change during the study year.

# Impact of COVID-19 on human mobility and retail sales in the US

By focusing on the yearly change in retail that occurred in April 2020, it is possible to identify the greatest impact of the pandemic. These losses, however, vary dramatically by state with Hawaii, Alaska, California, New Jersey, and New York (HI, AK, CA, NJ, and NY) being the most affected and Mississippi, Arkansas, Alabama, Tennessee, and Vermont (MS, AR, AL, TN, and VT) the least affected. A bar chart ranks the states according to these losses with the most significantly affected states highlighted (Figure 3). Hawaii experienced the largest shock because of the reduction in tourism with many people unable or unwilling to fly. WalletHub found Hawaii to be the hardest-hit state by COVID-19 due to its dependency on tourism (Sweet, 2020).

Table 1 lists the states that experienced the lowest and highest yearly change in retail, including the magnitude of the loss, and their respective region and division. This shows that the most resilient states are located in the south region. At the same time, the west and northeastern regions were severely affected because they are popular tourist destinations and are the most economically developed and are also densely populated. Voting behavior also appears to be associated with the impact. Hawaii, California, New Jersey, and New York were Democrat states based on the US 2020 Presidential Election, with Alaska the only Republican state in this group of most severely impacted states. In contrast, most of the resilient states are Republican's.

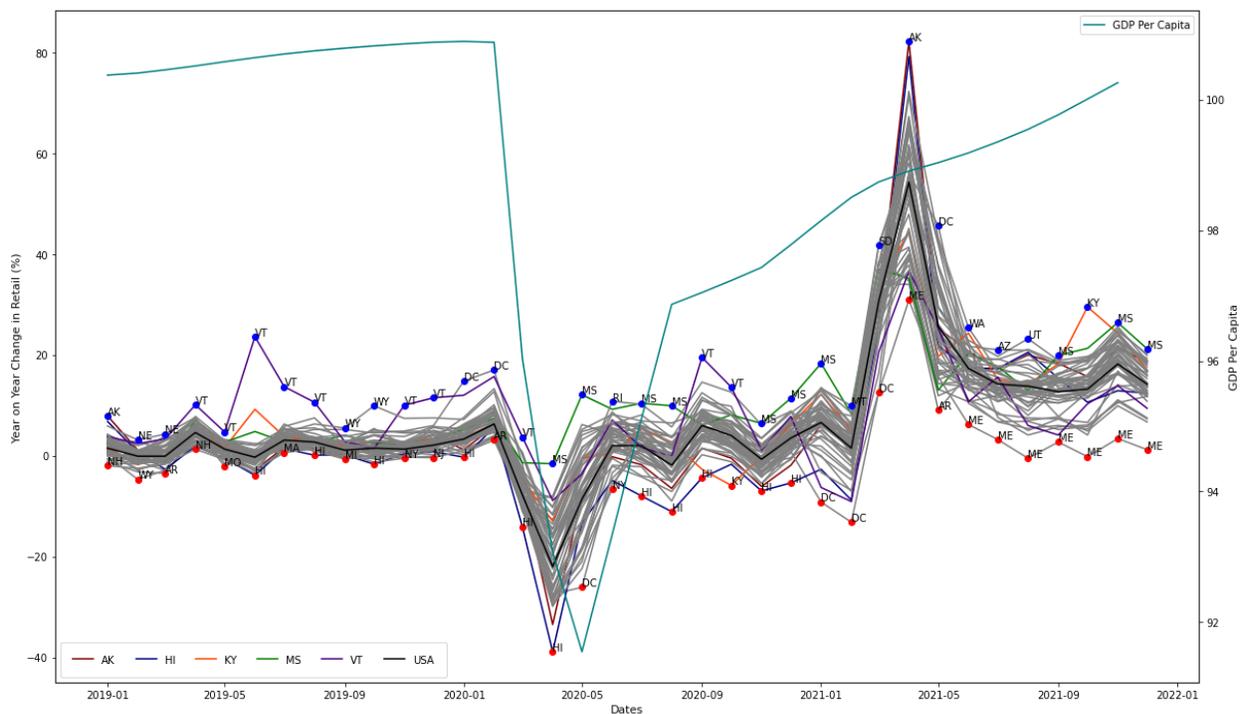

Figure 2: Yearly change in retail and GDP per capita for the US states from January 2019 to December 2021.

# Impact of COVID-19 on human mobility and retail sales in the US

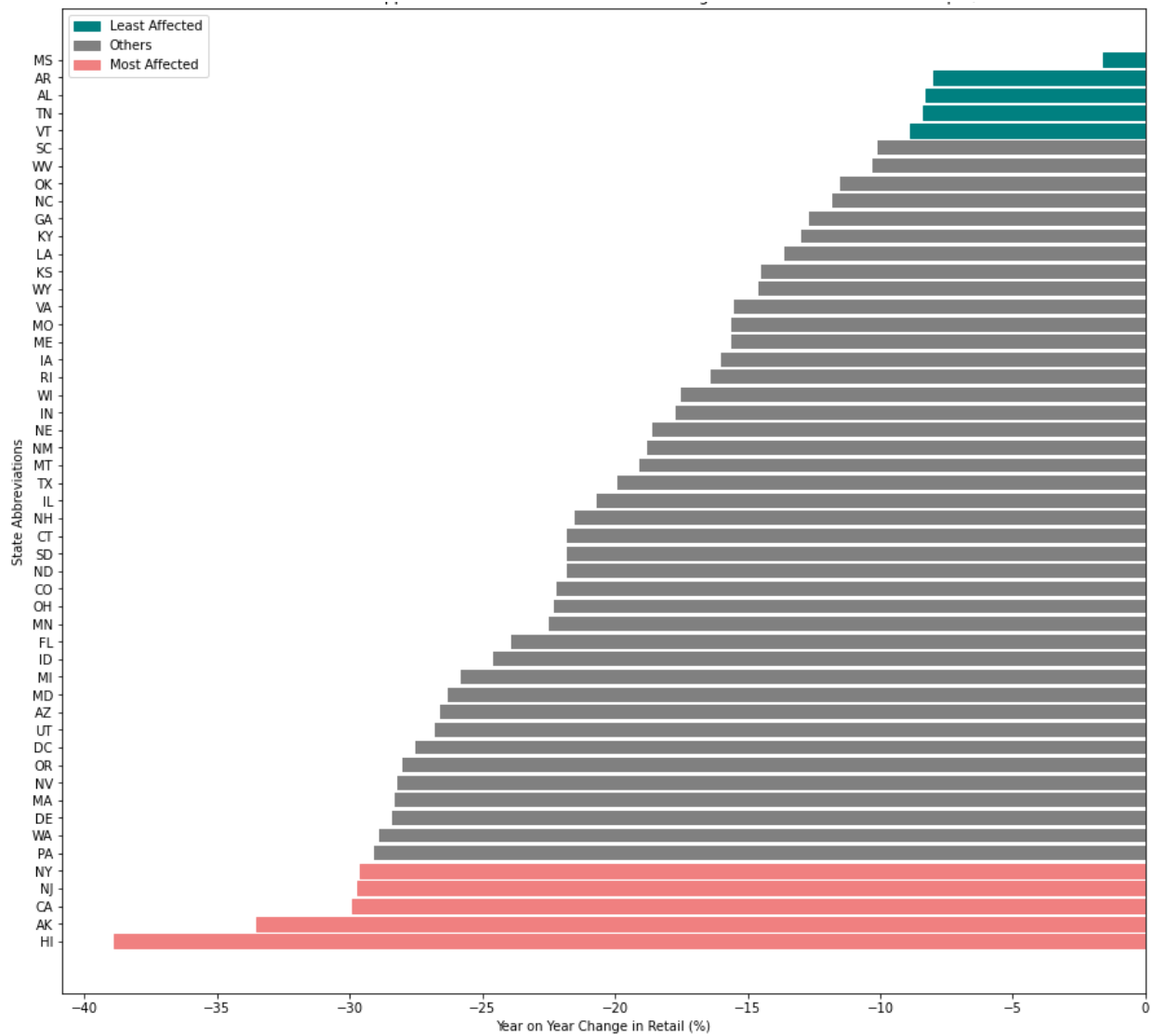

Figure 3: Barchart shows the yearly financial losses for retail sales in April 2020 for each US state.

Table 1: Summary states whose retail sales were least and most affected, showing yearly change in April 2020, region, and division.

| Five Least Affected States | | | | |
|---|---|---|---|---|
| State Name | State Abbreviation | Percentage | Region | Division |
| Mississippi | MS | -1.6 | South | East South Central |
| Arkansas | AR | -8.0 | South | West South Central |
| Alabama | AL | -8.3 | South | East South Central |

| Tennessee | TN | -8.4 | South | East South Central |
|---|---|---|---|---|
| Vermont | VT | -8.9 | NorthEast | New England |
| Five Most Affected States ||||| 
| Hawaii | HI | -38.9 | West | Pacific |
| Alaska | AK | -33.5 | West | Pacific |
| California | CA | -29.9 | West | Pacific |
| New Jersey | NJ | -29.7 | NorthEast | Middle Atlantic |
| New York | NY | -29.6 | NorthEast | Middle Atlantic |

**3.3 Identifying factors that influenced the size of the losses**

The period that sustained the most significant shock due to COVID-19 is the reported decline in retail sales for April 2020 (Figure 2). By taking these yearly changes in retail sales observed in April 2020 as the dependent variable, it is possible to consider different state-specific socio-economic characteristics that are likely to explain the large variability in losses between states (Figure 3). Note that the changes are negative for all states and that the dependent variable is therefore a collection of losses. This implies that variables with a positive correlation contribute to larger losses. The correlation of these losses with socio-economic explanatory variables serves to shed some light on the varying outcomes (Table 2). According to these correlations, the states that were called Democrat in the 2020 Presidential Election were burdened with the largest losses. There are other factors to consider, such as Hawaii which suffered a tremendous loss of -38.9% because the island is highly dependent on tourism. A ranking of US states in terms of tourism by WalletHub confirms that Hawaii is the leading state for tourism and was the worst hit by COVID-19 (Sweet, 2020).

Table 2: Correlation between state-level characteristics and retail losses in April 2020

| State Level Characteristics | Correlation | R - Squared (%) | Ranking |
|---|---|---|---|
| Percentage of Republican Vote | 0.501 | 25.1 | **1** |
| South Region | 0.491 | 24.1 | **2** |
| Republican | 0.486 | 23.6 | **3** |
| Percentage of Democrat Vote | -0.486 | 23.6 | **4** |
| Percent of adults with a high school diploma only | 0.477 | 22.7 | **5** |

# Impact of COVID-19 on human mobility and retail sales in the US

| Percent of adults with a bachelor's degree or higher | -0.456 | 20.8 | 6 |
|---|---|---|---|
| West Region | -0.456 | 20.8 | 7 |
| GDP Per Capita | -0.408 | 16.6 | 8 |
| Percent of adults with less than a high school diploma | 0.311 | 9.7 | 9 |
| State Population | -0.211 | 4.5 | 10 |
| Population Density | -0.150 | 2.3 | 11 |
| Region_Northeast | -0.131 | 1.7 | 12 |
| Median Household Income Per capita | -0.124 | 1.5 | 13 |
| Land Area | 0.092 | 0.9 | 14 |
| Midwest Region | 0.040 | 0.2 | 15 |
| Percent of adults completing some college or associate's degree | 0.025 | 0.1 | 16 |

Backward stepwise regression was used to select the features that best explain the variability in losses. Table 3 provides information about the regression model including estimated parameters, t-statistics, and p-values. The model explains 54.7% of the variability as indicated by the adjusted R-square of the linear regression used to fit the data. A negative coefficient indicates a factor that contributes to greater losses. An increase of 1% in the proportion of the US population residing in a given state is associated with an increase of 1.3% in annual retail loss reported in April 2020. Likewise, a 1.0% decrease in the percentage of adults with less than a high school diploma is associated with a 1.2% increase in retail loss. The more educated a state is, the more that state suffers from retail losses. A 10% increase in the proportion of people voting Democrat in a state is associated with a 2.4% increase in retail sales loss. To summarize, the regression reveals that the largest financial losses occurred in states with large populations, a high percentage of Democrat voters, and a low percentage of adults with less than a high school diploma (indicative of a highly educated population) and are most likely located in the west region.

Table 3: Regression of the yearly change in retail sales recorded in April 2020 on state-level characteristics.

| Selected Features | Coefficient | t-statistic | p-value |
|---|---|---|---|

# Impact of COVID-19 on human mobility and retail sales in the US

| Constant | -0.1634 | -3.319 | < 0.01 |
|---|---|---|---|
| State Population | -1.2875 | -3.177 | < 0.01 |
| Percentage of Democrat Vote | -0.2461 | -3.805 | < 0.001 |
| Percent of adults with less than a high school diploma | 1.2055 | 3.634 | < 0.01 |
| West Region | -0.0747 | -4.386 | < 0.001 |
| R-squared: | 0.583 | | |
| Adjusted R-squared | 0.547 | | |

### 3.4 Quantifying how mobility drives retail sales

The available information facilitates a comparison of the trip by distance daily average with yearly changes in retail sales, merged on the month and US state. The explanatory variables obtained from the Department of Transport are the number of people at home, the number of people not at home, and trip data defined by the distance of each trip. All the mobility explanatory variables have been normalized by population and are per capita. The yearly change in the retail sales for each month is the dependent variable. The average correlation across the 51 states is computed for each explanatory variable. The variable with the highest correlation is the very short trips of less than one mile, which is primarily responsible for driving retail sales in each US state. By aggregating the trips based on their durations, it was possible to construct aggregate variables for short, medium, and long trips. The influence on retail sales was found to consistently decline for longer trips as shown by the decreasing correlation values and regression coefficients (Table 4). This implies that very short trips are the strongest predictor of retail sales and an increase of very short trips by 1% is associated with a 0.46% increase in retail sales .

Table 4: Summary of average correlation and average regression between mobility and yearly change in retail across 51 US states for each trip variable from January 2019 to December 2021

| Trip Variables Per Capita | Correlation, % | Regression coefficient |
|---|---|---|
| Very short trip (trip less than 1 mile) | 50.9 | 0.4628 |
| Short trip (trips less than 10 miles) | 40.0 | 0.3959 |
| Medium trip (trips from 10 miles to | 28.1 | 0.3003 |

# Impact of COVID-19 on human mobility and retail sales in the US

| less than 100 miles) | | |
|---|---|---|
| Long trip (trips of 100 miles and more) | 15.7 | 0.1435 |

The correlation between very short trips and yearly change in retail sales is positive . This correlation varies across the 51 states (Figure 4) with the highest value of 80% in Iowa (IA) and the lowest of 10% in Maine (ME) .

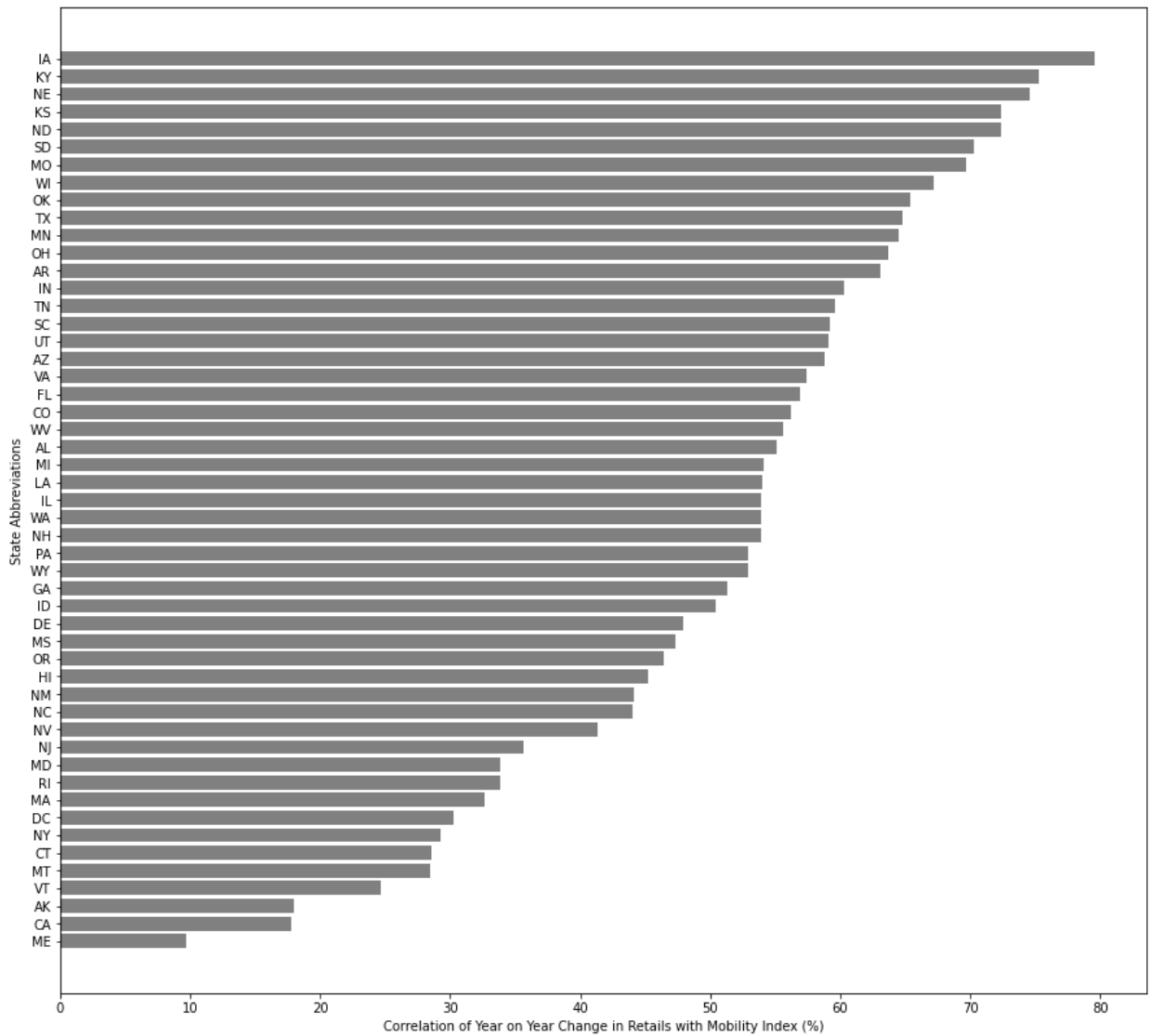

Figure 4: Correlation between mobility index and yearly change in retail across 51 US states.

**3.5 Compliance and dependence on mobility**

# Impact of COVID-19 on human mobility and retail sales in the US

COVID-19 resulted in government responses of varying stringency to restrict mobility in an attempt to prevent the spread of the virus. The US states experienced varying levels of stringency over time and the citizens responded in different ways. The behavior of individuals is best measured using the mobility data that demonstrates how many trips were taken. Finally, the reduction in mobility had the negative effect of reducing retail sales as customers were no longer able to make trips. The very short trips of less than one mile per capita is defined as the mobility index and were already shown to be the most highly correlated with retail sales. In this way, it is possible to study the entire causal chain from government policy (stringency) to mobility to retail sales.

To completely understand these linkages, it is therefore necessary to study two relationships. First, the correlation between stringency and mobility, as measured by mobility index, is calculated as a measure of compliance. Second, the correlation of mobility index and yearly change in retail sales is utilized to measure the dependence of the retail sector on mobility  We then determine what drives these patterns by again regressing on state-level explanatory variables such as education, income, population, presidential election voting patterns, and geographic regions.

### 3.6 Compliance

To measure compliance, the negative of the correlation between stringency and mobility index is calculated. This correlation measures how many trips are being made versus changes in the stringency index. If the inhabitants of a state are compliant, then it is expected that the correlation will be negative (fewer trips when the stringency index is high). The more negative the correlation, the more compliant the state. For each of the US states, the correlation is negative (Figure 5) which confirms overall compliance with the government rules (Figure 6). However, the more negative the value, the fewer trips that take place with increasing stringency, indicating that some states are more compliant than others. California, New Mexico, and Arizona were found to be the most compliant. In contrast, Oklahoma, Alabama, and Louisiana are the least compliant. It was also found that the Democrat states are more compliant (average of 53.1%) than the Republican states (average of 39.6%) based on the voting patterns in the 2020 Presidential Election which is inline with the finding from (Shvetsova et al., 2021). The dependency of compliance on state-level characteristics assessed using correlation suggests the differences are most influenced by political voting patterns, population, and education level (Table 5).

Table 5: Correlation of compliance with state-level characteristics.

| State Level Characteristics | Correlation | R - Squared(%) | Ranking |
|---|---:|---:|---:|
| Democrat Vote | -0.558 | 31.2 | 1 |
| Republican Vote | 0.556 | 30.9 | 2 |
| Republican | 0.545 | 29.7 | 3 |

# Impact of COVID-19 on human mobility and retail sales in the US

| | | | |
|---|---|---|---|
| State Population | -0.441 | 19.4 | **4** |
| Percent of adults with a high school diploma only | 0.378 | 14.3 | **5** |
| Percent of adults with a bachelor's degree or higher | -0.318 | 10.1 | **6** |
| Population Density | -0.302 | 9.1 | **7** |
| South Region | 0.254 | 6.4 | **8** |
| GDP Per Capita | -0.250 | 6.2 | **9** |
| West Region | -0.237 | 5.6 | **10** |
| Percent of adults completing some college or associate's degree | 0.223 | 5.0 | **11** |
| Percent of adults with less than a high school diploma | -0.160 | 2.6 | **12** |
| Median Household Income Per capita | 0.159 | 2.5 | **13** |
| Northeast Region | -0.139 | 1.9 | **14** |
| Midwest Region | 0.087 | 0.8 | **15** |
| Land Area | 0.060 | 0.4 | **16** |

Backward stepwise regression was used to select statistically significant features that serve to explain the compliance level of each state. The result of the regression reveals the percentage of Democrat votes, state population, and being situated in the west region are statistically significant variables that explain compliance. Table 6 provides the model summary in terms of coefficients, t-statistics, and P-values. The resultant model explains 45.8% of the variability in compliance as indicated by the adjusted R-square (Table 6). The result shows a 10% increase in the proportion of the US population living in a state will yield a 2% increase in compliance. A 10% rise in the proportion of people voting Democrat is associated with a 5% increase in compliance. In summary the result suggests that states with a high population, a high percentage of Democrat voters, and those located in the western region tend to be more compliant.

Impact of COVID-19 on human mobility and retail sales in the US

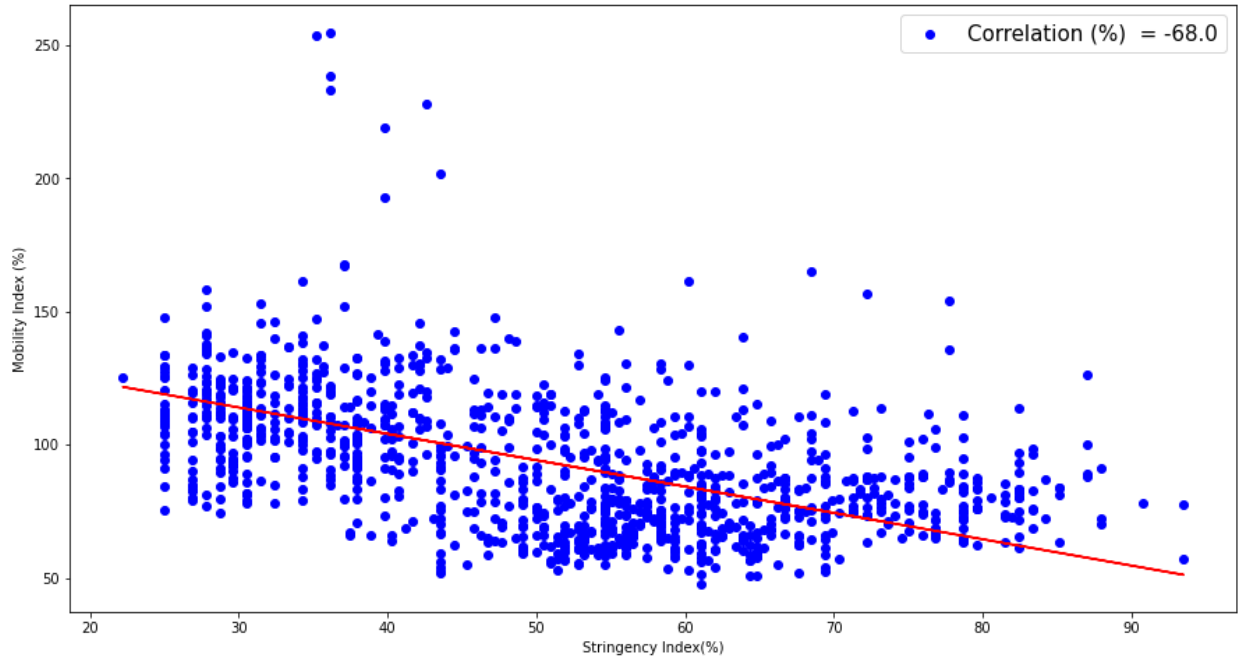

Figure 5: Scatterplot of the mobility index against the stringency index.

# Impact of COVID-19 on human mobility and retail sales in the US

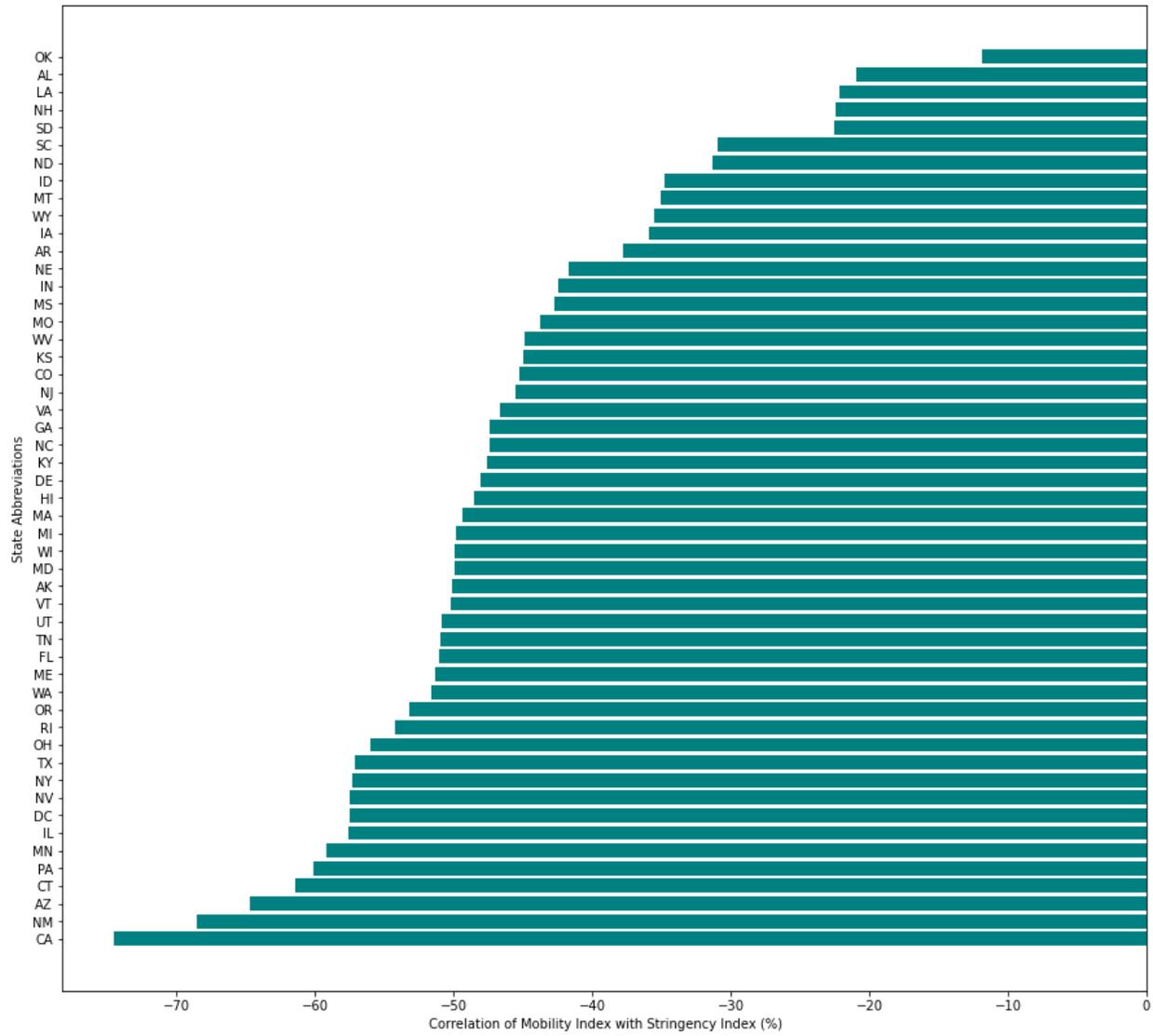

Figure 6: Variation in compliance across states, defined as the negative correlation between the stringency index and mobility index.

Table 6: Regression of compliance on state-level characteristics.

| Selected Features | coefficient | T- statistic | P-value |
| --- | --- | --- | --- |
| Constant | 0.1590 | 2.893 | < 0.01 |
| Population Proportion | 1.9927 | 3.340 | < 0.01 |
| Democrat Vote | 0.5112 | 4.617 | < 0.001 |

| West Region | 0.0718 | 2.418 | < 0.05 |
|---|---|---|---|
| R-squared | 0.490 | | |
| Adjusted R-squared | 0.458 | | |

### 3.7 Dependence of retail on mobility

Yearly changes in retail sales are highly correlated with mobility (very short trips) and the previous results confirmed that, as expected, more trips generated more sales (Figure 4). These correlations, one value for each state, are taken as a dependent variable and regressed on state-level characteristics. The sensitivity of the retail sector to mobility using correlation suggests the differences are influenced by geography, level of education, and voting patterns (Table 7). Backward stepwise regression was again used to select the variables that explain the dependence of retail on mobility. The model result identified the Midwest, Northeast, and West region, and percentage of Democrat vote as the statistically significant variables (Table 8). A key result is that an increase of one percent in the Democrat vote is associated with a decrease of half a percent in the dependence of retail on mobility. The final model explains 58.3% of the variability according to the adjusted R-squared and demonstrates that there is a higher dependence of retail on mobility in states with a low percentage of democrat voters and located in the Midwest region while the dependence is lower for states in the Northeast and West regions (Table 8).

Table 7: Correlation of dependence of retail on mobility with state-level characteristics.

| State Level Characteristics | Correlation | R - Squared(%) | Ranking |
|---|---|---|---|
| Republican Vote | 0.582 | 33.9 | **1** |
| Democrat Vote | -0.567 | 32.2 | **2** |
| Midwest Region | 0.549 | 30.1 | **3** |
| Northeast Region | -0.502 | 25.2 | **4** |
| Republican | 0.484 | 23.5 | **5** |
| Percent of adults with a bachelor's degree or higher | -0.408 | 16.6 | **6** |
| Percent of adults completing some college or associate's degree | 0.360 | 12.9 | **7** |
| Median Household Income Per capita | -0.291 | 8.5 | **8** |
| Land Area | 0.289 | 8.4 | **9** |
| GDP Per Capita | -0.284 | 8.0 | **10** |
| Percent of adults with a high school diploma only | 0.264 | 7.0 | **11** |
| West Region | -0.250 | 6.2 | **12** |
| Population Density | -0.193 | 3.7 | **13** |
| South Region | 0.143 | 2.0 | **14** |
| Resident Population 2020 Census | -0.099 | 1.0 | **15** |
| Percent of adults with less than a high school diploma | 0.026 | 0.1 | **16** |

Table 8: Regression of dependence of retail on mobility (very short trips) on state-level characteristics.

| Selected Features | Coefficient | T- statistic | P-value |
|---|---|---|---|
| Constant | 0.7782 | 11.495 | < 0.001 |
| Democrat Vote | -0.5059 | -3.768 | < 0.001 |
| Midwest Region | 0.1142 | 2.976 | < 0.01 |
| Northeast Region | -0.1489 | -2.239 | < 0.01 |
| West Region | -0.0920 | -2.374 | < 0.05 |
| R-squared: | 0.617 | | |
| Adjusted R-squared: | 0.583 | | |

## 4. Discussion

The COVID-19 pandemic led to the use of lockdowns and restrictions on human mobility being implemented across the world. The aim of these restrictions was to prevent the spread of the

# Impact of COVID-19 on human mobility and retail sales in the US

virus by suppressing the transmission as individuals were unable to socialize. While the public health benefits were easy to justify with mobility restrictions rolled out in the majority of countries around the world, there have been many debates about the economic losses caused by these interventions. With hindsight, it is now possible to explore the relationships between the varying government policy responses to the pandemic, the behavior of the population when faced with mobility restrictions and the consequences for the economy.

Calculations of the stringency of these government response policies by OxCGRT facilitate a detailed analysis of how individuals respond to rules and guidelines with changes in mobility patterns—essentially allowing direct measurement of compliance at the state level. Compliance was defined as the negative correlation between stringency level and mobility index.

Furthermore, monthly records of the yearly change in retail sales for each US state provide an ideal opportunity to study the economic impact of these policies that were designed to suppress the spread of the virus and effectively manage COVID-19. By using trip data from the USBTS, it was possible to identify the key mobility factors that determine retail sales.

This study has successfully answered all three research questions: (1) to quantify the financial loss of retail due to COVID-19; (2) to measure variability in impact by the US states; and (3) to explain the difference in effects using the explanatory variables. In the US, government response policies generated substantial financial losses for the retail sector. Annual US GDP fell by -3.5% in 2020, the largest contraction since 1946. The economy contracted by as much as -31.2% in the second quarter due to the lockdowns. Mobility is a crucial factor for understanding retail sales. The COVID-19 pandemic that led to government policies aimed at suppressing transmission and saving lives also had a drastic effect on retail sales. By requesting people to stay at home and restricting movement, the number of fatalities was reduced. US retail sales declined by -22% in April 2020 compared to the previous year. These yearly losses registered in April 2020 varied dramatically by state, ranging from -1.6% in Mississippi to -38.9% in Hawaii. A regression of the yearly change in retail sales recorded in April 2020 on state-level characteristics was able to explain 54.7% of the variability in these losses. The greatest financial losses occurred in states with large populations, a high percentage of Democrat voters, a low percentage of adults with less than a high school diploma, and located in the West region.

Geographical location is a critical factor, with states in the West and Northeast regions suffering the largest retail losses while the South region was relatively resilient. Very short trips of less than one mile per capita are, on average, most strongly correlated with retail sales. In general, the longer the trip distance the less impact there is on retail sales. An increase of 1% in the proportion of the US population residing in a given state is associated with an increase of 1.3% in annual retail loss reported in April 2020. Likewise, a 1.0% decrease in the percentage of adults with less than a high school diploma is associated with a 1.2% increase in retail loss. The more educated a state's population, the more that state suffered from retail losses. In addition a 10% increase in the proportion of people voting Democrat in a state is associated with a 2.4% increase in retail sales loss.

Compliance was quantified by measuring the correlation between the stringency index and the mobility index. The more negative the correlation, the more compliant the state. While all states were on the whole compliant, variations were further analyzed using state-level characteristics. States that voted Democrat in the 2020 Presidential Election were the most compliant with an average compliance of 53.1% compared to 39.6% for Republican states. However, the largest retail losses tended to occur in Democrat states rather than Republican states. Very short trips of less than one mile per capita defined as the mobility index had the greatest influence on retail sales, achieving a correlation of 50.9% whereas longer trips were less correlated with retail sales. This finding suggests that the longer the trip, the more likely it is for purposes other than retail. This makes sense given that many people tend to shop near their homes and that longer trips are reserved for commuting or vacations. It was found that a 10% increase in the mobility index is associated with a 4.6% increase in retail sales.

States with a large population, a high number of Democratic voters, a low percentage of adults with less than a diploma certificate, and located in the West region are more likely to be compliant. A 10% increase in the proportion of the US population living in a state is associated with a 2% increase in compliance. Furthermore, a 10% rise in the proportion of people voting Democrat is associated with a 5% increase in compliance.

Retail sales in states with a low percentage of Democrat voters, located in the Midwest, Northeast and Western region tend to have a high dependence on mobility. States in the West region were most affected and those in the South were the most resilient. The study also revealed that states that voted democrat in the 2020 Presidential Election were more compliant. These Democrat states, however, also suffered the largest retail losses. Similarly, an increase of one percent in the Democrat vote is associated with a decrease of half a percent in the dependence of retail on mobility. This also suggests that Republican states tend to show a stronger reliance on mobility to drive retail.

## 5.0 Conclusions

The purpose of this study is to investigate the impact of COVID-19 on the retail sector at the state-level in the US. While government response policies successfully curtailed the spread of the disease and reduced fatalities, the economic cost of doing so has been substantial.

A deeper understanding of the adverse effects on the retail sector at the spatial level of each state was possible due to having monthly data on the stringency of the policies, several mobility measures in terms of trip distances, and the yearly changes in retail sales. There are many differences between the states, and socio-economic factors such as education, income, election voting patterns, population, area, GDP, and region were considered in an attempt to explain the relationships between policies, mobility and retail sales. The key findings obtained from this study are as follows:
- The largest retail losses tended to occur in Democrat states.
- States in the West region were most affected and those in the South were resilient.
- An increase of 1% in the proportion of the US population residing in a given state is associated with an increase of 1.3% in annual retail loss reported in April 2020.
- Mobility is a statistically significant driver of retail sales and the correlation between them decreases with longer trip distances.
- A mobility index defined as the number of very short trips of less than one mile per capita is most strongly correlated with retail sales.
- A 10% increase in this mobility index is associated with a 4.6% increase in retail sales.
- Democrat states tend to be more compliant and a 10% rise in the proportion of people voting Democrat is associated with a 5% increase in compliance.
- A 10% increase in the proportion of the US population living in a state is associated with a 2% increase in compliance.
- States with a low percentage of adults with less than a diploma certificate are more likely to be compliant.
- An increase of 10% in the stringency index is associated with a decrease of 10% in the mobility index
- Republican states are most dependent on mobility to drive retail sales.

The datasets acquired during the pandemic offer a unique opportunity to better understand the compliance levels of individuals when reacting to government policies and also how mobility patterns drive retail sales. It is hoped that policymakers, practitioners, and researchers can use this work to formulate appropriate policies to mitigate the adverse effects of future pandemics while ensuring the economy does not suffer.

## Limitation
The available data supported the calculation of compliance based on mobility and stringency. While there have been many policies that have been implemented by the government to cope with the pandemic, it is important to stress that non-mobility policies such as promoting the use of masks is not covered by this research. Furthermore, this study focused on the total value of all retail sales and did not distinguish between different types of retail firms.

## Recommendations
We recommend striking a careful balance when using mobility restrictions to suppress the spread of disease. This study has shown mobility, specifically trips of less than one mile, is the strongest predictor of retail sales and that limiting longer trips of greater than one mile is preferable to avoid the negative consequences of disruption to the retail sector. Additionally, higher education levels appear to be associated with greater compliance and therefore a focus on education is beneficial.

The availability of subnational retail sales, mobility data, voting patterns, education, and population information allowed this project to study many relationships. It is recommended that other countries collect these valuable datasets, thereby facilitating valuable insights, improving decision-making and preparing for future pandemics.